\documentclass[sigconf]{acmart}
\renewcommand\footnotetextcopyrightpermission[1]{} % removes footnote with conference information in first column
\AtBeginDocument{%
	\providecommand\BibTeX{{%
			\normalfont B\kern-0.5em{\scshape i\kern-0.25em b}\kern-0.8em\TeX}}}
\newtheorem{definition}{Definition}[section]
\newtheorem{Proposition}{Proposition}
\usepackage{ amssymb   }
\usepackage{ mathdots  }
\usepackage{ amsmath   }
\usepackage{ color     }
\usepackage{ graphicx  }
\usepackage{ amsfonts  }
\usepackage{ amssymb   }
\usepackage{ epstopdf  }
\usepackage{ upgreek   }
\usepackage{ bm        }
\usepackage{ graphicx  }
\usepackage{ subfigure }
\usepackage{ mathtools }
\usepackage{ algorithm }
\usepackage{algorithmic}
\usepackage{ booktabs  }
\usepackage{ multirow  }
\usepackage{ caption   }
\usepackage{ array     }
\usepackage{ float     }
\usepackage{   soul    }
\usepackage[normalem]{ulem}
\newcommand{\norm}[1]{\left\lVert#1\right\rVert}

\begin{document}
	% ===========================
	%          title
	% ===========================
	\title{Single-Layer Graph Convolutional Networks For Recommendation}
	
	% ===========================
	%          author
	% ===========================
	% author
	\author{Yue Xu}
	\affiliation{\institution{Beijing University of Posts and Telecommunications}}
	\email{xuy@bupt.edu.cn}
	
	% author
	\author{Hao Chen}
	\affiliation{\institution{Beihang University}}
	\email{chh@buaa.edu.cn}
	
	% author
	\author{Zengde Deng}
	\affiliation{\institution{The Chinese University of Hong Kong}}
	\email{zddeng@se.cuhk.edu.hk}
	
	% author
	\author{Junxiong Zhu}
	\affiliation{\institution{Alibaba Group}}
	\email{xike.zjx@taobao.com}
	
	% author
	\author{Yanghua Li}
	\affiliation{\institution{Alibaba Group}}
	\email{yichen.lyh@taobao.com}
	
	% author
	\author{Peng He}
	\affiliation{\institution{WeChat, Tencent Inc}}
	\email{paulhe@tencent.com}
	
	% author
	\author{Wenyao Gao}
	\affiliation{\institution{WeChat, Tencent Inc}}
	\email{tankygao@tencent.com	}
	
	% author
	\author{Wenjun Xu}
	\affiliation{\institution{Beijing University of Posts and Telecommunications	}}
	\email{wjxu@bupt.edu.cn}
	\renewcommand{\shortauthors}{}
	
	% ============================
	%          abstract
	% ============================
	\begin{abstract}
		Graph Convolutional Networks (GCNs) and their variants have received significant attention and achieved start-of-the-art performances on various recommendation tasks. However, many existing GCN models tend to perform recursive aggregations among all related nodes, which arises severe computational burden. Moreover, they favor multi-layer architectures in conjunction with complicated modeling techniques. Though effective, the excessive amount of model parameters largely hinder their applications in real-world recommender systems. 
		To this end, in this paper, we propose the single-layer GCN model which is able to achieve superior performance along with remarkably less complexity compared with existing models. Our main contribution is three-fold.
		First, we propose a principled similarity metric named distribution-aware similarity (DA similarity), which can guide the neighbor sampling process and evaluate the quality of the input graph explicitly. We also prove that DA similarity has a positive correlation with the final performance, through both theoretical analysis and empirical simulations. 
		Second, we propose a simplified GCN architecture which employs a single GCN layer to aggregate information from the neighbors filtered by DA similarity, and then generates the node representations. Moreover, the aggregation step is a parameter-free operation, such that it can be done in a pre-processing manner to further reduce red the training and inference costs. 
		%、
		Third, we conduct extensive experiments on four datasets. The results verify that the proposed model outperforms existing GCN models considerably and yields up to a few orders of magnitude speedup in training, in terms of the recommendation performance. 
	\end{abstract}
	
	% ============================
	%        Keywords
	% ============================
%	\keywords{Graph Convolutional Networks, Recommendation, Heterogeneous Information Network}
	
	\fancyhead{}
	\settopmatter{printacmref=false,printfolios=false}
	
	\maketitle
	
	% ============================
	%        Introduction
	% ============================
	\section{Introduction}
	Recommender system plays a pivotal role in various online services, e.g., E-commerce, news feeds, and video-on-demand services. The aim of recommendation is to match user preference with resource items~\cite{zhang2019deep}. 
	Traditional recommendation models, e.g., matrix factorization~\cite{koren2009matrix} and collaborative filtering~\cite{sarwar2001item}, mainly model user preference by performing statistical analysis on historical user-item interaction records. 
	Nowadays, as various kinds of auxiliary data become increasingly available in online services, many recommendation models shift their focus to graph-based methods~\cite{zhao2017meta,ying2018graph, zhao2019intentgc, fan2019meirec, wu2019dual, wang2019kgat, wang2019knowledge, Wang2019graph}, which have greater expressive power on modeling manifold types of nodes and relationships in recommender systems. 
	
	Among others, Graph Convolutional Networks (GCNs), which generalize the Convolutional Neural Networks~(CNNs) on graph-structured data~\cite{kipf2016semi}, have achieved impressive performance on various graph-based learning tasks~\cite{velivckovic2017gat,hamilton2017representation,hamilton2017inductive}, including recommendation~\cite{zhao2019intentgc}. The core idea behind GCNs is to iteratively aggregate information from locally nearby neighbors in a graph using neural networks~\cite{chen2018fastgcn}. Specifically, each node at one GCN layer performs graph convolution operation to aggregate information from its nearby neighbors at the previous layer. By stacking multiple GCN layers, the information can be propagated across far reaches of a graph, which makes GCNs capable of learning from both content information as well as graph structure. As such, GCN-based models are widely adopted in recommendation tasks~\cite{ying2018graph,zhao2019intentgc, fan2019meirec, wu2019dual, wang2019kgat, wang2019knowledge, Wang2019graph} which require learning from relational datasets. 
	%Recently, GCN models have been extended to learn from heterogeneous graph for recommendation, aiming at modeling both explicit and implicit interactions between user and items. For example, ...
	However, although existing GCN-based recommendation models have set a new standard on many benchmark tasks~\cite{ying2018graph,zhao2019intentgc, fan2019meirec, wu2019dual, wang2019kgat, wang2019knowledge, Wang2019graph}, they suffer from two main pitfalls.
	
	\smallskip\noindent\textbf{Recursive Neighborhood Aggregation.}
	The recursive neighborhood aggregation among all nodes arises severe computational burden, which, however, may have limited contribution in recommendation tasks.
	Specifically, as pointed out in~\cite{li2018laplacian}, the convolution in GCN model is indeed a special form of Laplacian smoothing, which mixes the features of a node and its nearby neighbors. The smoothing operation makes the feature of nodes within the same cluster to be similar, thus greatly easing the classification/regression task.
	Therefore, \textit{it is critical for GCN models to ensure that similar nodes have been grouped into the same cluster before performing the aggregations.}
	In homogeneous networks, it is highly likely for two similar nodes to form a direct edge in the graph, which is known as the homophily hypothesis~\cite{mcpherson2001homophily}. In this case, by recursively aggregating features from $1$-hop neighbors, GCNs are able to achieve impressive performances~\cite{kipf2016semi,hamilton2017inductive,huang2018asgcn}. 
	
	However, in the context of recommendation in heterogeneous networks, the difficulty of recognizing similar nodes arises since we need to measure the similarity between two users (or items) based on their \textit{indirect} relationships. 
	In particular, existing models usually measure the similarity between two users (or items) according to their historical interactions with other auxiliary nodes.
	For example, \cite{fan2019meirec, Wang2019graph, wu2019dual, wang2019knowledge} consider two users to be similar if they clicked the same item or the same brand, which, however, can be easily dominated by the popular items or brand; \cite{zhao2019intentgc} measures the similarity of two users according to the number of their common auxiliary neighbors. However, in this case, the users who interacted with most of the auxiliary nodes would have a high similarity to all other users. Besides, the number of common neighbors is unlikely to scale linearly with the value of similarity.
	Additionally, none of them defines an explicit and principled metric to quantitatively evaluate the node similarity in heterogeneous networks. 
	In fact, given such a similarity metric, we may not need to perform recursive aggregations with multiple GCN layers. Instead, we only need to select similar neighbors for each node beforehand, and then perform aggregation for only once with a single GCN layer. 
	
	\smallskip\noindent\textbf{Complicated Architecture.}
	Many existing models suffer from considerable computational complexity due to the use of multi-layer architectures in conjunction with complicated modeling techniques.
	For example, the metapath-guided GCN models~\cite{hu2018leveraging, fan2019meirec} construct manifold metapaths to find similar neighbors for aggregations, which arises more complexity on both information aggregation and data pre-processing. The attention based GCN~(GAT) models~\cite{velivckovic2017gat, wang2019kgat, wu2019dual} generalize the graph convolution with the attention mechanism, which, however, introduce additional and excessive amount of model parameters. Besides, \cite{wang2019hgat} further introduce a contextual multi-arm bandit over GAT to weight the interactions of various social effects, which brings higher uncertainties in model tuning.
	Generally, to some extent, these models are trading complexity for potential performance enhancement, which largely hinder their application in real-world recommender systems. 
	
	On the other hand, the recent advances on simplified GCNs such as~\cite{wu2019sgc}, indicate that it is feasible to remove certain components from existing architectures while still preserving comparable performances. 
	This motivates us to rethink about the essential components of building an expressive GCN model for recommendations. Moreover, exploring the existence of an efficient and effective GCN architecture is not only a must for the application to current recommender platforms, but also paves the way for the resources-constrained on-device~(e.g., mobile phone and wearable devices) recommendation in the near future.
	
	\smallskip\noindent\textbf{Our Work.} In this paper, we consider the user-item recommendation problem and propose the single-layer GCN~(SLGCN) model. The model has a much lower complexity compared to existing GCN-based recommendation models but is able to achieve superior performance. The main contributions are summarized as follows.
	\begin{itemize}
		\item \textbf{Principled Similarity Metric:} we propose a principled similarity metric named distribution-aware similarity~(DA similarity) which explicitly measures the similarity of a pair of nodes according to the distribution of their interactions towards other auxiliary nodes. On this basis, we propose another quantitative metric named Mean Average Neighbor Similarity~(MANS) to evaluate the quality of neighbor sampling results. Then, we prove that MANS has a positive correlation with the final recommendation performance from a theoretical standpoint.
		Experimental results verify our analysis and show that existing GCN models can also benefit from our proposed similarity metric to improve the performance, without changing their model architectures. 
		
		\item \textbf{Simplified Learning Architecture:} we propose a simplified GCN architecture which generates node representations with only a single GCN layer. Particularly, the architecture performs propagation for only once to aggregate information from the neighbors which are selected based on DA similarity.
		Moreover, the aggregation step is indeed a parameter-free operation, such that it can be done in a pre-processing manner to further reduce the training costs. 
		Besides, we also investigate the efficiency of different  architectures of the prediction layer.
		
		\item \textbf{Extensive Verifications:} we conduct extensive experiments on three benchmark datasets and one commercial dataset to verify the superiority of our proposed model. The results show that our proposed model can outperform existing GCN models considerably, and yield up to a few orders of magnitude speedup in training.
	\end{itemize}
	
	% ============================
	%        Related Work
	% ============================
	\section{Related Work}
	\subsection{GCN-based Recommendation}
	GCNs originated from a version of graph convolutions developed based on spectral graph theory~\cite{kipf2016semi} and have many variants on various fields, e.g., node classification~\cite{hamilton2017inductive,velivckovic2017gat,chen2018fastgcn}, link prediction~\cite{zhang2018linkprediction,chami2019hgcn}, as well as recommendation~\cite{zhao2017meta,ying2018graph,zhao2019intentgc}.
	The user-item recommendation aims at directly predicting users' preference over items. Related GCN models usually first generate user and item embeddings by utilizing both content information and graph structure, and then predict user-item interactions~\cite{zhao2019intentgc,fan2019meirec, hu2018leveraging, fan2019graph}.
	While most models adopt multiple multi-layer perception~(MLP) layers to construct the prediction layer, their architectures to obtain node representation differ from each other.
	In particular, IntentGC~\cite{zhao2019intentgc} proposed the vector-wise convolution to avoid useless feature interactions during neighborhood feature propagation. MEIRec~\cite{fan2019meirec} leveraged LSTM to capture the sequential correlation among different neighbors. 
	KGAT~\cite{wang2019kgat} computed the hidden states of each node by attending over its neighbors. 
	Dual Graph Attention Networks~\cite{wu2019dual} introduced a contextual multi-arm bandit to weight social influence on the user's preference for items.
	However, all these models are constructed with a stack of multiple nonlinear GCN layers, which requires fitting excessive amount of model parameters.
	On the other hand, the recently proposed simple graph convolution (SGC)~\cite{wu2019sgc} reveals that removing certain components (the nonlinear transformations in their work) from GCNs causes little effect on the performance of node classification. This encourages us to seek for a compact but effective model architecture in the context of recommendation.

	\subsection{Similarity Measurement}
	Existing recommendation models proposed various strategies to measure the node similarities which are then used to guide the neighbor sampling process.
	Among others, the most popular strategy is based on the first-order proximity. In particular, many models consider two users (or two items) to be similar if they have interacted with the same auxiliary node. The sampling probability can either depend on the interaction frequency~(i.e., importance sampling) or not (i.e., random sampling). Examples include MEIRec~\cite{fan2019meirec}, KGCN~\cite{Wang2019graph}, Dual Graph Attention Networks~\cite{wu2019dual}, KGNN-LS~\cite{wang2019knowledge}, etc.
	The other choice is based on the second-order proximity, which measures the similarity of two nodes by comparing their neighborhood structure~\cite{goyal2018graph}. For example, IntentGC~\cite{zhao2019intentgc} measured the similarity between two nodes by comparing the number of their common neighbors.
	Another group of works such as Pinsage~\cite{ying2018graph} leveraged the random walk to measure the similarity.
	However, all these works only provide empirical explanations on  similarity measurements, without developing an explicit similarity metric or investigating the influence of similarity measurement (or neighbor sampling) on final recommendation performance.
	Besides, there are also recent works from other fields studied the graph sampling methods~\cite{zeng2020graphsaint,li2018laplacian, chen2018fastgcn, huang2018asgcn}.
	The most related work to ours is LINE~\cite{tang2015line} which proposed to measure nodes' similarity by comparing their distributions. However, they defined the distribution from a perspective of generating network context, which is different from ours. Besides, their aim is to propose an optimization objective for network embedding, while we aim at GCN-based recommendation.
	
	% ============================
	%          Section
	% ============================
	\section{Problem Definition}
	In this paper, we consider the user-item recommendation task within a graph consists of heterogeneous nodes and relationships. 
	Specifically, the user-item recommendation task can be described as follows.
	We denote the user set as $ \mathcal{U} = \{ u_1, u_2, \cdots, u_N\} $ with $ N $ the number of users, and denote the item set as $ \mathcal{I} = \{ i_1, i_2, \cdots, i_M \} $ with $ M $ the number of items. Given a user node $ u \in \mathcal{U}$ and an item node $ i \in \mathcal{I} $, the aim of user-item recommendation is to predict the potential interaction $ r_{u, i} $ (e.g., click, rate, and purchase) between user $ u $ and item $ i $. 
	On the other hand, the heterogeneous graph can be modeled as a heterogeneous information network~(HIN), which is defined as follows:
	\begin{definition}[Heterogeneous Information Network]
		\label{def:HIN}
		A HIN is defined as a graph $\mathcal{G}=(\mathcal{V}, \mathcal{E})$ where $\mathcal{V}$ is the set of nodes and $\mathcal{E}$ is the set of edges between the nodes in $\mathcal{V}$. Each node $v \in \mathcal{V}$ and each edge $ e \in \mathcal{E}$ is associated with a node type mapping function $ \phi: \mathcal{V} \rightarrow \mathcal{T}_v $ and an edge type mapping function $ \varphi: \mathcal{E} \rightarrow \mathcal{T}_e $, respectively. The number of types satisfy $ |\mathcal{T}_v|>1 $ or $ |\mathcal{T}_e|>1 $.
	\end{definition}

	Moreover, we consider the GCN models are trained with a subgraph sampled from the entire graph. This is a practical setting in real-world recommender systems~\cite{zhao2019intentgc, fan2019meirec, ying2018graph}, since training GCN models with the entire graph $\mathcal{G}=(\mathcal{V}, \mathcal{E})$ will arise excessive computational complexity.
	Specifically, we consider each node in the graph only aggregates information from a subset of its neighbors. The sampled subgraph can be represented as $\mathcal{G}^{sub}=(\mathcal{V}, \mathcal{E}^{sub})$, where $ \mathcal{E}^{sub} \subseteq \mathcal{E} $ denotes the edges between each node and its sampled neighbors. Note that the subgraph still contains the entire set of nodes $ \mathcal{V} $ from the original graph $ \mathcal{G} $ (where $ \mathcal{U}, \mathcal{I} \in \mathcal{V} $), but only contains a subset of edges (i.e., propagation paths among the nodes) from the original graph due to neighbor sampling. 
	In other words, the sampling process only reduces the information aggregated from the neighbors, without removing any node from the graph. In this case, it is critical to sample the most similar neighbors for each node in $ \mathcal{G}^{sub} $ in order to guarantee reliable performance.
	
	We aim to 
	1) propose a principled and interpretable similarity metric to guide the neighbor sampling process and investigate the influence of neighbor sampling on the recommendation performance;
	2) propose an efficient and effective GCN architecture which is able to achieve superior performance to existing models but with much lower complexity.
	
	% ============================
	%          Section
	% ============================
	\section{Neighbor Sampling}
	\subsection{Network Translation}
	Recommendation models mainly focus on modeling user nodes and item nodes. Therefore, it is a common routine for them to translate all relationships in the original graph into user-user and item-item relationships~\cite{zhao2019intentgc, fan2019meirec, ying2018graph, fan2019graph}. In this way, they can avoid modeling all different types of nodes and relationships, thereby reducing the model complexity. 
	In the translated graph, two users (or items) are considered to have one connected path if they have both interacted with the same auxiliary node. For example, two users are considered to be connected if they clicked the same item or purchased the same brand. 
	The subgraph is constructed by allowing each node in the translated graph to sample its neighbors according to their inter-connected paths.
	
\begin{figure}[t]
		\centering
		\subfigure[First-order Proximity]
		{ 	
			\label{fig:ord1}
			\includegraphics[trim = 10 20 10 10, clip, width=0.3\columnwidth]{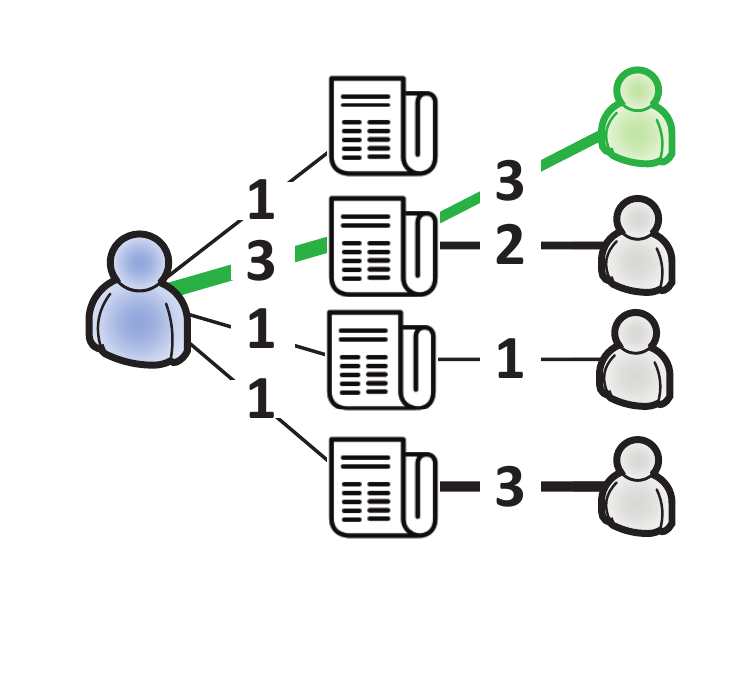}
		}
		\subfigure[Second-order Proximity]
		{ 	
			\label{fig:ord2}
			\includegraphics[trim = 10 20 10 10, clip, width=0.3\columnwidth]{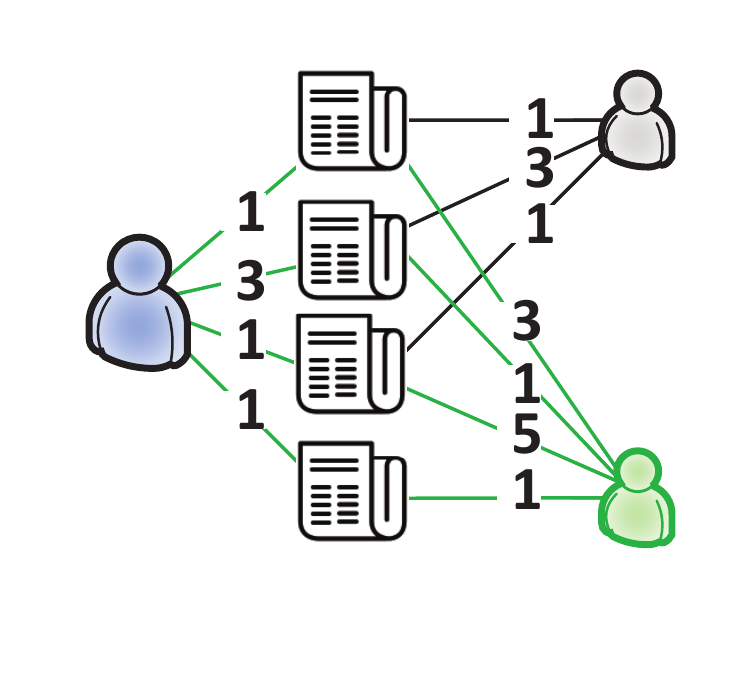}
		}
		\subfigure[DA Similarity]
		{ 	
			\label{fig:DA}
			\includegraphics[trim = 10 20 10 10, clip, width=0.3\columnwidth]{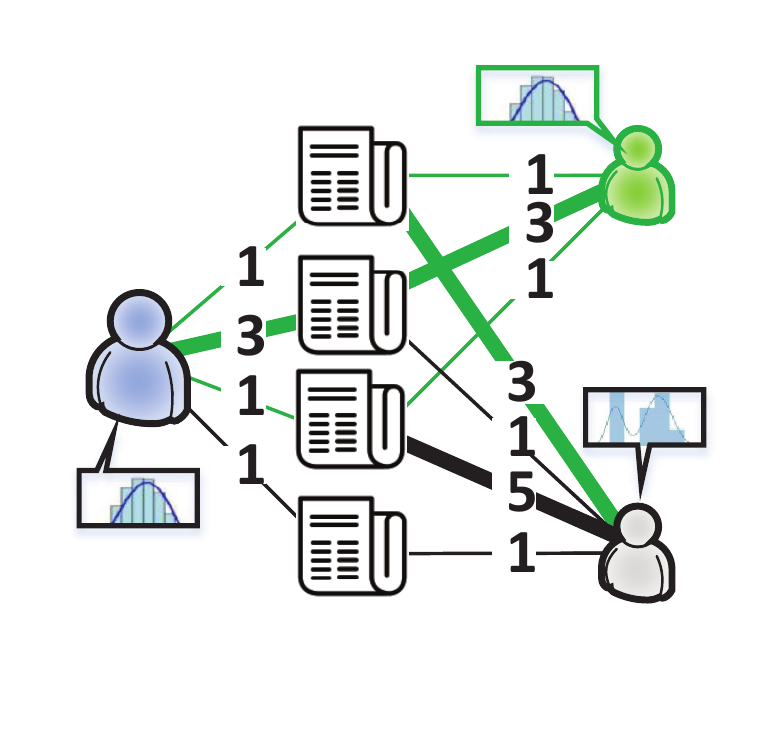}
		}
		\vskip -1em
		\caption{Examples of neighbor sampling with different similarity metrics. (a) Sampling according to the weights of direct edges. (b) Sampling according to the number of common item-clicks. (c) Sampling according to the distribution of item-clicks.}
		\vskip -1em
		\label{fig:performance}
	\end{figure}

	Existing works proposed various similarity metrics to guide the sampling process.
	Among others, the first-order proximity and the second-order proximity are the most popular ones.
	Specifically, the first-order proximity measures the similarity between two nodes according to the weight of their connected path~\cite{goyal2018graph}. 
	Taking user-click-item paths as an example, as shown in Figure~\ref{fig:ord1}, the target user finds its $ 2 $-hop neighbors by first traversing to his/her top clicked items (1-hop), and then traversing to the item's top clicked users (2-hop). The traversing probability can either depend on the path weight (i.e., importance sampling) or not (i.e., random sampling). However, this method can be easily influenced by the popular nodes whose paths usually have higher weights than the others.
	Alternatively, the second-order proximity measures the similarity between two nodes according to the proximity of their neighborhood structure~\cite{goyal2018graph}.
	For example, as shown in Figure~\ref{fig:ord2}, the target user measures the similarity of each neighbor according to the number of their common item-clicks. However, in this case, the users who clicked most of the items would have a high similarity towards all other users. Also, the number of common neighbors is unlikely to scale linearly with the value of similarity. Inspired by the above methods, we next propose a more principled similarity metric that takes both path weights and neighborhood structure into consideration.
	
	\subsection{Distribution-Aware Similarity}
	We propose the DA similarity in the context of recommendation, which measures the similarity between two nodes according to their interaction distribution upon other nodes. 
	
	For clarity, let us first consider the user-click-item paths. We denote a user $ u_n \in \mathcal{U}$'s click probability over an item $ i_m \in \mathcal{I} $ as $ p_{u_n}(i_m) $ and denote his/her click probability over all items in $ \mathcal{I} $ as $ \mathbf{P}_{u_n}(\mathcal{I}) $. Then, the similarity between user $ u_1 $ and user $ u_2 $ on item-click preference can be written as $ d\big(\mathbf{P}_{u_1}(\mathcal{I}), \mathbf{P}_{u_2}(\mathcal{I}) \big) $. 
	This similarity can be formulated with various distance metrics defined on probability distribution. For example, with the Kullback-Leibler divergence~(KL divergence), the distance between user $ u_1 $'s and user $ u_2 $'s preference on item-clicks can be formulated as 
	\begin{equation}\label{key}
	d_{KL}(\mathbf{P}_{u_1},\mathbf{P}_{u_2}) = \sum\nolimits_{i_m \in \mathcal{I}^+_{u_1} \cup \mathcal{I}^+_{u_2} } p_{u_1}(i_m) \ln \frac{p_{u_1}(i_m)}{p_{u_2}(i_m)},
	\end{equation}
	where $ \mathcal{I}^+_{u_1} $ refers to the set of user $ u_1 $'s clicked items. 
	We define the similarity as the negative distance between $ \mathbf{P}_{u_1} $ and $ \mathbf{P}_{u_2} $, i.e.,
	\begin{equation}\label{key}
	\xi_{KL}(\mathbf{P}_{u_1},\mathbf{P}_{u_2}) = -~d_{KL}(\mathbf{P}_{u_1},\mathbf{P}_{u_2}).
	\end{equation}
	As such, higher similarity means less distance on the probability distribution.
	The similarity formulated by KL divergence is asymmetric, we can also formulate a symmetric DA similarity with the norm function:
	\begin{equation}\label{key}
	\xi~(\mathbf{P}_{u_1}, \mathbf{P}_{u_2}) = -~d_{\rm norm}~(\mathbf{P}_{u_1}, \mathbf{P}_{u_2}) =  - \norm{\mathbf{P}_{u_1} - \mathbf{P}_{u_2}},
	\end{equation}
	where $ \norm{\cdot} $ denotes the norm function, including the L1-norm $ \norm{\cdot}_1 $ function and the L2-norm $ \norm{\cdot}_2 $ function, etc.
	
	\smallskip\noindent\textbf{Heterogeneous relationships.}
	Now we extend the definition of DA similarity under heterogeneous relationships. Considering a graph with $ |\mathcal{T}_v| $ types of nodes and $ |\mathcal{T}_e| $ types of relationships, the definition of DA similarity can be given as follows.
	\begin{definition} [Distribution-Aware Similarity]
		\label{def:similarity}
		Given a graph $\mathcal{G}=(\mathcal{V}, \mathcal{E})$ with $ |\mathcal{T}_v| $ types of node and $ |\mathcal{T}_e| $ types of relationships, we define the set of probability distributions of node $x$'s interaction with other nodes under all relationships as
		\begin{equation*}
		\mathbf{P}_{x}=\{\mathbf{P}_{x}^{v,e}\}_{v\in\mathcal{T}_v,e\in\mathcal{T}_e},
		\end{equation*}
		where $\mathbf{P}_{x}^{v,e}$ is a probability distribution denoting the probability of node $x$ to interact with other nodes of type $ v $ under the relationship of type $ e $. 
		Then, the DA similarity between node $ x $ and node $ y $ can be written as
		\begin{equation}
		\xi \big(\mathbf{P}_{x}, \mathbf{P}_{y} \big) =  - \sum_{e=1}^{|\mathcal{T}_e|} \lambda_e \sum_{v=1}^{|\mathcal{T}_v|} \lambda_v \cdot d\big(\mathbf{P}_{x}^{v,e}, \mathbf{P}_{y}^{v,e} \big),
		\end{equation}
		where $d(\cdot)$ is a distance function while $ \lambda_v $ and $ \lambda_e $ are the importance weights assigned to the similarity of interactions with the auxiliary nodes of type $ v $ under the relationship of type $ e $, respectively.
	\end{definition}
	
	\subsection{Neighbor Quality Measurement}
	\label{sec:quality}
	The DA similarity provides an explicit metric to measure the quality of neighbors. As such, the subgraph can be constructed by letting each node in the graph sample its neighbors according to the evaluated distribution-aware similarities. For example, the node could 1) directly select the top-$ K $ neighbors with the highest similarity or 2) normalize all neighbors' similarities into a probability distribution and perform importance sampling.
	
	Now we investigate the correlation between our defined similarity metric and the final prediction performance. To this end, we first give another quantitative metric named Mean Average Neighborhood Similarity~(MANS) to evaluate the quality of the sampled neighbors and 2) reveal the positive correlation between MANS and the prediction performance through theoretical analysis. 
	
	First, we define the average neighbor similarity~(ANS) as follows. 
	\begin{definition}[Average Neighborhood Similarity]
		\label{def:mans}
		For a given node $ u $, its average neighborhood similarity~(ANS) is defined as
		\begin{equation}\label{key}
		\Xi~(u) = \frac{1}{|\mathcal{N}_u|}\sum_{ v \in \mathcal{N}_u} \xi~(\mathbf{P}_u, \mathbf{P}_v),
		\end{equation}
		where $ \mathcal{N}_u $ is the set of sampled neighbors of node $ u $ and $ |\cdot| $ denotes the cardinality of a set. 
	\end{definition}
	ANS is the average DA similarity of one node's sampled neighbors, which measures the quality of one node's sampled neighbors. On this basis, the definition of mean average neighbor similarity~(MANS) of a sampled subgraph is given as follows. 
	
	\begin{definition}[Mean Average Neighbor Similarity]
		For a given sampled subgraph $ \mathcal{G}^{sub} $, its mean average neighborhood similarity~(MANS) is defined as 
		\begin{equation}\label{key}
		\Xi~(\mathcal{G}^{sub}) = \frac{1}{|\mathcal{V}|} \sum_{v \in \mathcal{V}} \Xi(v),
		\end{equation}
		where $ \mathcal{V} $ denotes the set of all the nodes in the subgraph.
	\end{definition}
	MANS is the mean of all nodes' ANS values in the subgraph, such that higher MANS indicates that the grouped nodes (each node and its neighbors) have a higher similarity. 
	Considering that the philosophy behind GCN is to smooth features over similar vertices thus easing the classification task~\cite{li2018laplacian}, it is highly likely that MANS has a positive correlation with the performance achieved by GCN models.
	Recall that in GCN models, the node embedding generated by the $l$-th layer can be generally written as
	\begin{equation}\label{equ:agg}
	\mathbf{h}_{u}^{(l)} =  \sigma\Big( \mathbf{W} \cdot \text{AGGREGATE}_{l} \left\lbrace \mathbf{h}^{(l-1)}_i, i\in{\mathcal{N}^+_u} \right\rbrace \Big), ~~\forall u \in \mathcal{V},
	\end{equation}
	where  $ \text{AGGREGATE}_{l} (\cdot) $ denotes the aggregation function at the $l$-th layer, $ \mathbf{W} $ refers to the linear transformation, $ \sigma(\cdot) $ is a nonlinear activation function, and $ \mathcal{N}^+_u $ is the union set of node $ u $ and its neighbors. 
	As pointed out in SGC~\cite{wu2019sgc}, the nonlinearity transformation between consecutive GCN layers can be redundant, since the main benefits of aggregation come from local averaging.
	Therefore, in order to highlight the influence of neighbor selection, we herein develop our theoretical analysis based on the SGC~\cite{wu2019sgc}. 
	In this case, the update function in~(\ref{equ:agg}) can be simplified into
	\begin{equation}
	\label{eq:sgc_agg}
	\textbf{h}_u^{(l)} = \frac{1}{|{\mathcal{N}^+_u}|}\sum_{v\in{\mathcal{N}^+_u}}  \mathbf{h}_{v}^{(l-1)}, ~~\forall u \in \mathcal{V}.
	\end{equation}
	
	Next, for clarity and ease of derivation, we analyze the aggregation process on user-click-item paths as an example.
	User modeling aims at generating an accurate user embedding to describe his/her preference on the item-click event. 
	Given any user $ u \in \mathcal{U} $, we denote the estimated probability distribution on his/her item-click event as $\tilde{\mathbf{P}}^{(l)}_u = F(\mathbf{h}_u^{(l)})$, where $ \mathbf{h}_u^{(l)} $ is his/her embedding generated by the $l$-th layer and $F(\cdot)$ is an unbiased mapping function. 
	In order to generate an accurate item-click prediction, the GCN model needs to minimize the distance between the true probability distribution $ \mathbf{P}_u $ and the estimated probability distribution $ \tilde{\mathbf{P}}_u $. The distance can be measured by the KL divergence:
	\begin{equation}\label{eq:distance}
	d_{KL}\big(\mathbf{P}_u,\tilde{\mathbf{P}}^{(l)}_u \big) = \sum_{i \in \mathcal{I}} P_u(i) \ln \frac{P_u(i)}{\tilde{P}^{(l)}_u(i)}.
	\end{equation}
	Given a node $ v_i \in {\mathcal{N}^+_u} $, we denote the estimated probability distribution on his/her item-click event by the $ (l\!-\!1) $-th layer as $ \tilde{\mathbf{Q}}^{(l-1)}_{v_i} $. Without loss of generality, we assume that the probability distributions of different neighbors are independent from each other, such that we have $ \tilde{\mathbf{Q}}(u) = \frac{1}{|{\mathcal{N}^+_u}|} \sum_{v_i \in {\mathcal{N}^+_u}} \tilde{\mathbf{Q}}^{(l-1)}_{v_i}  $ from~(\ref{eq:sgc_agg}).
	Therefore, the KL distance given in~(\ref{eq:distance})  satisfies
	\begin{subequations}\label{eq:kl-proof}
		\begin{align} 
		&d_{KL}\big(\mathbf{P}_u,\tilde{\mathbf{P}}^{(l)}_u\big) = \sum_{i \in \mathcal{I}} P_u(i)  \ln \frac{P_u(i) }{\tilde{P}^{(l)}_u(i)}\\
		& = \sum_{i \in \mathcal{I}} P_u(i) \ln \frac{P_u(i)}{\frac{1}{|{\mathcal{N}^+_u}|} \sum_{v_i \in {\mathcal{N}^+_u}} \tilde{Q}^{(l-1)}_{v_i}(i)} \\
		& \leq\sum_{i \in \mathcal{I}} P_u(i) \Bigg( \ln P_u(i) - \frac{1}{|{\mathcal{N}^+_u}|} \sum_{v_i \in {\mathcal{N}^+_u}} \ln \tilde{Q}^{(l-1)}_{v_i}(i)\Bigg) \label{eq:js}\\
		& = \frac{1 }{|{\mathcal{N}^+_u}|} \sum_{v_i \in {\mathcal{N}^+_u}} \sum_{i \in \mathcal{I}} P_u(i) \Big(\ln P_u(i) - \ln \tilde{Q}^{(l-1)}_{v_i}(i)\Big) \\
		& = \frac{1 }{|{\mathcal{N}^+_u}|} \sum_{v_i \in {\mathcal{N}^+_u}} d_{KL}\big(\mathbf{P}_u,\tilde{\mathbf{Q}}^{(l-1)}_{v_i}\big),
		\end{align}
	\end{subequations}
	where the inequality in (\ref{eq:js}) is based on the Jenson inequality~\cite{boyd2004convex}. The results in~(\ref{eq:kl-proof}) reveals that the the distance between $ \mathbf{P}_u $ and $ \tilde{\mathbf{P}}_u $ is upper bounded by the distance between $ \mathbf{P}_u $ and all $ \tilde{\mathbf{Q}}^{(l-1)}_{v_i} $, i.e., 
	\begin{equation}\label{eq:KLupperbound}
	d_{KL}(\mathbf{P}_u,\tilde{\mathbf{P}}^{(l)}_u) \leq \frac{1}{|{\mathcal{N}^+_u}|} \sum_{v_i \in {\mathcal{N}^+_u}} d_{KL}(\mathbf{P}_u,\tilde{\mathbf{Q}}^{(l-1)}_{v_i}),
	\end{equation}
	where $ d_{KL}(\mathbf{P}_u,\tilde{\mathbf{Q}}^{(l-1)}_{v_i}) $ is an approximation of $ d_{KL}(\mathbf{P}_u,\mathbf{Q}_{v_i}) $.
	Hence, one can minimize the upper bound of $ d_{KL}(\mathbf{P}_u,\tilde{\mathbf{P}}^{(l)}_u) $ by minimizing $ \frac{1}{|{\mathcal{N}^+_u}|} \sum_{v_i \in {\mathcal{N}^+_u}} d_{KL}(P_u,Q_{v_i}) $, which means increasing the ANS value of node $ u $. 
	In other words, the probability of correctly predicting user's item-click interaction (i.e, estimating $ \tilde{P}_u $) can be increased by sampling his/her neighbors with higher DA similarity values. 
	Note that the result in (\ref{eq:kl-proof}) also holds when formulating the distance with norm functions, which satisfies
	\begin{subequations}
		\begin{align} \label{eq:norm-proof}
		&d_{\text{norm}}\big(\mathbf{P}_u, \tilde{\mathbf{P}}^{(l)}_u \big) = \norm{\mathbf{P}_u - \frac{1}{|{\mathcal{N}^+_u}|} \sum\nolimits_{v_i \in {\mathcal{N}^+_u}} \tilde{\mathbf{Q}}^{(l-1)}_{v_i}}\\
		& = \frac{1}{|{\mathcal{N}^+_u}|} \norm{\sum\nolimits_{v_i \in {\mathcal{N}^+_u}} \left( \mathbf{P}_u - \tilde{\mathbf{Q}}^{(l-1)}_{v_i} \right)}\\
		& \leq \frac{1}{|{\mathcal{N}^+_u}|} \sum\nolimits_{v_i \in {\mathcal{N}^+_u}}\norm{ \mathbf{P}_u - \tilde{\mathbf{Q}}^{(l-1)}_{v_i} } \label{eq:triangle}\\
		& = \frac{1 }{|{\mathcal{N}^+_u}|} \sum\nolimits_{v_i \in {\mathcal{N}^+_u}} d_{\text{norm}}\big(\mathbf{P}_u,\tilde{\mathbf{Q}}^{(l-1)}_{v_i}\big),
		\end{align}
	\end{subequations}
	where the inequality in (\ref{eq:triangle})  comes from triangle inequality.
	
	The above analysis can be readily applied to other types of nodes and relationships in the heterogeneous graph. We therefore give the following proposition.
	\begin{Proposition}\label{the:proposition}
		When learning from a sampled subgraph $ \mathcal{G}^{sub} \subseteq \mathcal{G} $, it is promising to increase the performance of GCN-based recommendation models by increasing MANS of the subgraph, i.e., increasing $ \Xi~(\mathcal{G}^{sub}) = \frac{1}{|\mathcal{V}|} \sum_{u \in \mathcal{V}} \Xi~(u) $.
	\end{Proposition}
	
	% ============================
	%          Section
	% ============================
	\section{Single-layer GCN}
	In this section, we propose an efficient single-layer GCN architecture to learn the node representation. The architecture performs propagation for only once to aggregate information from the neighbors which are selected based on DA similarity, without suffering from the excessive computation caused by recursive aggregations. 
	Moreover, the aggregation step in our architecture is indeed a parameter-free operation which can be done in a pre-processing manner, thus can significantly reduce the model complexity as well as training and inference costs.
	
	\begin{figure*}[tb]
		\label{fig:error}
		\centering
		\includegraphics[trim = 20 10 10 10, clip, width=1.65\columnwidth]{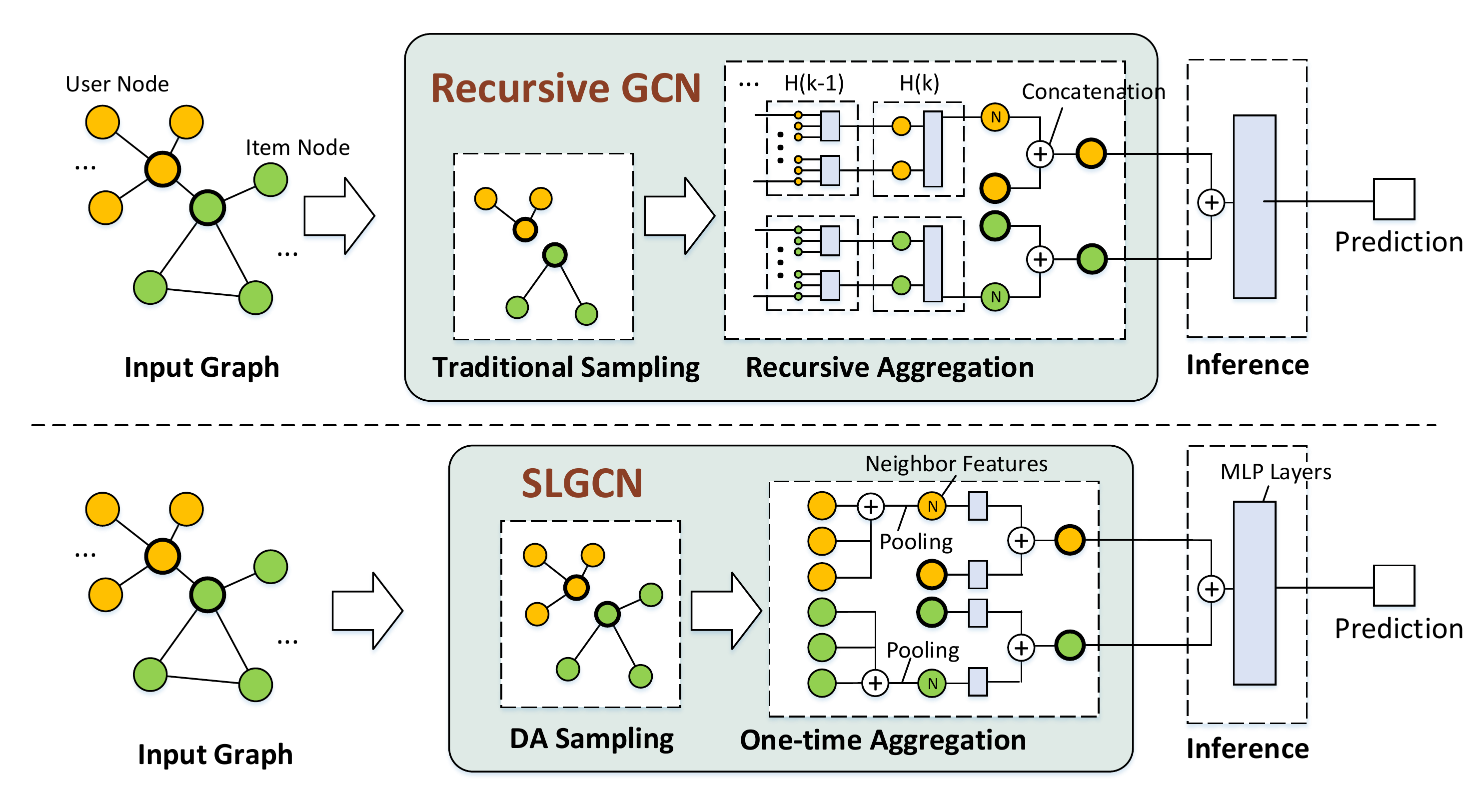}
		\vskip -0.5em
		\caption{The architectures of traditional recursive GCNs v.s. our proposed SLGCN. Top panel: the recursive GCN repeatedly performs propagations throughout $ K $ GCN layers. Bottom panel: the SLGCN performs propagation for only once among the neighbors filtered by the DA similarity metric.}
		\vskip -1em
	\end{figure*}
	\subsection{Node Representation}
	The user modeling and item modeling are symmetric in our proposed single-layer GCN architecture. Therefore, we mainly present user modeling for illustration in the following statement.
	Specifically, given a user $ u \in \mathcal{U} $, we initialize his/her embedding vector with the raw features, i.e., $ \mathbf{H}_u^{(0)} = \mathbf{X}_u $, where $ \mathbf{X}_u $ denotes the raw feature vector of user $ u $.
	Then, we aggregate the features from the neighbors of user $ u $ as the neighborhood feature, i.e.,
	\begin{equation}\label{eq:user-aggregation}
	\mathbf{X}_{\mathcal{N}_u} =  \text{AGGREGATE}\big\{ \mathbf{X}_v,~~\forall v \in \mathcal{N}_u \big\},
	\end{equation}
	where $ \mathcal{N}_u $ is the set of neighbors sampled according to the DA similarity, $ \mathbf{X}_v $ denotes the raw feature vector of the neighbor $ v \in \mathcal{N}_u $, and $ \text{AGGREGATE}\{ \cdot \} $ is a pooling function, e.g., mean pooling. 
	Afterwards, we generate the aggregated feature of user $ u $ by concatenating its self-feature and the neighborhood feature together, i.e., 
	\begin{equation}\label{eq:user-concatenation}
	\overline{\mathbf{X}}_u = \mathbf{X}_u \oplus \mathbf{X}_{\mathcal{N}_u},
	\end{equation}
	where $ \oplus $ denotes the concatenation of vectors. 
	Then, we feed the aggregated feature vector $ \overline{\mathbf{X}}_u $ into a \textit{single} neural network layer to obtain the user representation:
	\begin{equation}\label{eq:user-representation}
	\mathbf{H}_u = \sigma\Big( \mathbf{W} \cdot \overline{\mathbf{X}}_u \Big).
	\end{equation}
	Similarly, the item modeling is performed under the same process but needs to replace the context with item-related neighbors and features. 
	
	\smallskip\noindent\textbf{Remark.}
	It is noteworthy that both (\ref{eq:user-aggregation}) and (\ref{eq:user-concatenation}) are parameter-free operations since they do not require fitting any weights. As such, they are essentially equivalent to a feature pre-processing step. 
	In this case, the user/item modeling reduces to~(\ref{eq:user-representation}), which is only a simple transformation based on a single-layer neural network. 
	
	\smallskip\noindent\textbf{Heterogeneous Relationships.}
	We now extend SLGCN to deal with more heterogeneous relationships, which includes two methods.
	First, we could use the heterogeneous similarity metric $ \xi \big(\mathbf{P}_{x}, \mathbf{P}_{y} \big) $ in Definition~\ref{def:similarity} to select neighbors and follow the process from (\ref{eq:user-aggregation}) to (\ref{eq:user-representation}) to generate node representations. In this case, one needs to specify the hyperparameters $ \lambda_e $ and $ \lambda_v $ based on domain-knowledge, which is encouraged when dealing with familiar recommendation context. 
	Alternatively, we could put the hyperparameters into the concatenation step, i.e., (\ref{eq:user-concatenation}), to automatically determine the weights of different types of nodes and relationships in training. 
	For example, when considering the node type $ v_i $ with relationship type $ e_j $, the concatenation step can be modified into
	\begin{equation}\label{eq:user-concatenation-hete}
	\overline{\mathbf{X}}_u = \mathbf{X}_u \oplus \Big(\lambda_{e_i, v_j} \cdot \mathbf{X}_{\mathcal{N}^{e_i,\!v_j}_u}\Big),
	\end{equation}
	where $ \mathcal{N}^{e_i,\!v_j}_u $ denotes the set of similar neighbors filtered with the similarity metric $ d\big(\mathbf{P}_{x}^{e_i,\!v_j}, \mathbf{P}_{y}^{e_i,\!v_j} \big) $ and $ \lambda_{e_i,\!v_j} $ is the importance weight.
	
	\subsection{Prediction} 
	In this paper, we model the user-item recommendation task as a binary classification problem, where the positive label refers to an observed user-item interaction, and the negative label otherwise. 
	The prediction process can be formulated as
	\begin{equation}
	\label{eq:slgcn}
	\hat{y}_{u,i} = f (\mathbf{H}_u \oplus \mathbf{H}_i),
	\end{equation}
	where $\mathbf{H}_i$ is the item representation and $f(\cdot)$ denotes a mapping function. 
	The function $f(\cdot)$ can be constructed with a few MLP layers or with a dot product function. We will compare the performance of different choices in Sec.~\ref{sec:architecture}.
	We adopt the cross-entropy loss as our optimization objective, which can be given as
	\begin{equation}\label{eq:loss}
	J = \sum\nolimits_{u,i \in \mathcal{D}} \Big( y_{u,i} \log \hat{y}_{u,i} + (1-y_{u,i}) \log (1-\hat{y}_{u,i}) \Big),
	\end{equation}
	where $ \mathcal{D} $ denotes the training dataset, $ y_{u,i} $ is the real user-item recommendation label (equals $ 1 $ or $ 0 $), and $ \hat{y}_{u,i} $ is the predicted label.
	
	\subsection{Complexity Analysis} 
	\label{sec:complexity}
	The time cost of SLGCN mainly comes from a) subgraph construction, b) representation learning, and c) model inference.
	For a), we can offline compute the similarities of all connected users and items and then sample the neighbors for each node to construct the subgraph. Specifically, computing the similarity of a given user-user pair can be done in $ \mathcal{O}(K_{u'}+K_u) $ offline time, where $ K_u $ and $ K_{u'} $ denotes the nonzero interactions from the user to all items and from the other user to all items, respectively.
	Note that we only need to update the similarity matrix daily or weekly in practical recommender systems. 
	For b), we denote the complexity of performing pooling-based feature aggregation in (\ref{eq:user-aggregation}) to be $\mathcal{O}_{agg}$  
	and denote the complexity of representation mapping with MLP in (\ref{eq:user-representation}) to be $\mathcal{O}_{map}$. Without loss of generality, we assume that $\mathcal{O}_{agg}$ and $\mathcal{O}_{map}$ only differs with constant coefficient in different GCN models. 
	We denote the number of total training epochs as $E$, the number of total edges in the training set as $N_{train}$.
	The recursive GCN models~(\emph{e.g.} PinSAGE, MEIRec, IntentGC) perform recursive aggregations per training step. Moreover, they need to use MLP functions to do feature mapping after each aggregation step at each layer. We denote the number of total MLPs within the multiple GCN layers as $ L $. The complexity of recursive GCN models is $E \cdot N_{train}\cdot \mathcal{O}_{agg} + E \cdot N_{train}\cdot L \cdot  \mathcal{O}_{map}$. 
	Comparatively, SLGCN performs the aggregation in (\ref{eq:user-representation}) for only once during the pre-processing step, and performs feature mapping also for only once. As such, the complexity of SLGCN is $(M + N) \cdot \mathcal{O}_{agg} + E \cdot N_{train}\cdot \mathcal{O}_{map}$. Note that $ (M + N) \cdot \mathcal{O}_{agg} \ll E \cdot N_{train}\cdot \mathcal{O}_{agg} $.
	For c), we denote the number of prediction attempts as $N_{pred}$. The inference complexity of recursive GCNs is $N_{pred}\cdot \mathcal{O}_{agg} + N_{pred}\cdot \mathcal{O}_{map}$. While the inference complexity of SLGCN is only $N_{pred}\cdot \mathcal{O}_{map}$, since the neighbor aggregations have been completed beforehand. Empirical comparisons of the time costs of SLGCN vs other GCN models are presented in Sec.~\ref{sec:efficiency}.

	% ============================
	%          Section
	% ============================
		\label{sec:result}
	\begin{table*}[tbp]
		\small
		\centering
		\caption{Performance comparison on the four datasets.}
		\vskip -1em
		\setlength{\textwidth}{2.7mm}{
			\begin{tabular}{c|cc|cc|cc|cc}
				\toprule
				& \multicolumn{2}{c|}{\textbf{LastFM}} & \multicolumn{2}{c|}{\textbf{Ciao}} & \multicolumn{2}{c|}{\textbf{Epinions}} & \multicolumn{2}{c}{\textbf{WeChat}} \\
				\midrule
				& AUC   & NDCG@10  & AUC   & NDCG@10  & AUC   & NDCG@10  & AUC   & NDCG@10 \\
				\midrule
				MEIRec & 0.8723* & 0.7167* & 0.7705 & 0.5534 & 0.8363 & 0.7277 & 0.8036 & 0.6571 \\
				MEIRec++ & 0.8868 (+1.7\%) & 0.7167 (+0.0\%) & 0.8314 (+7.9\%) & 0.6289 (+13.6\%) & 0.8872 (+6.1\%) & 0.7985 (+9.7\%) & 0.9073 (+12.9\%) & 0.7343 (+11.7\%) \\
				\midrule
				IntentGC & 0.8704 & 0.7157 & 0.8123* & 0.6419* & 0.8574* & 0.7720* & 0.8808* & 0.7026* \\
				IntentGC++ & 0.8805 (+1.2\%) & 0.6826 (-4.6\%) & 0.8444 (+4.0\%)  & 0.6462(+0.6\%) & 0.8808 (+2.7\%) & 0.7766 (+0.6\%) & 0.9073 (+3.0\%) & 0.7345 (+4.5\%) \\
				\midrule
				SLGCN-1ord & 0.9348 & 0.7856 & 0.8656 & 0.7199 & 0.9003 & 0.8067 & 0.8574 & 0.5550 \\
				SLGCN-2ord & 0.9374 & 0.7871 & 0.8929 & 0.7628 & 0.9198 & 0.8125 & 0.9016 & 0.7411 \\
				SLGCN-sim2 & \textbf{0.9528} & \textbf{0.8112} & \textbf{0.9282} & \textbf{0.7957} & \textbf{0.9403} & \textbf{0.8280} & \textbf{0.9104} & \textbf{0.7602} \\
				\midrule
				
				Improvement & 9.2\% & 13.2\% & 14.3\% & 24.0\% & 9.6\% & 7.3\% & 3.4\% & 8.2\% \\
				\bottomrule
			\end{tabular}%
		}
		\vskip -1em
		\label{tab:result}
	\end{table*}
	\section{Experiments}
	We conduct extensive experiments on four datasets with the goal of answering four research questions:
	
	\smallskip\noindent\textbf{Q1:} Does our proposed SLGCN outperform the state-of-the-art GCN-based recommendation methods?
	
	\smallskip\noindent\textbf{Q2:} How efficient is the learning of SLGCN compared with other GCN-based architectures?
	
	\smallskip\noindent\textbf{Q3:} How does neighbor sampling affect the final performance?
	
	\smallskip\noindent\textbf{Q4:} What is the efficiency of different architectures for inference?
	
	\subsection{Experimental Setup}
	\smallskip\noindent\textbf{Datasets.}
	We use the following four datasets in our experiments for music, movie, products, and information recommendations, respectively:
	(1) \textbf{Last-FM}\footnote{https://grouplens.org/datasets/hetrec-2011/.} is \textcolor{black}{a music listening dataset collected from the Last.fm online music system}, where the tracks are viewed as items; 
	(2) \textbf{Ciao} \footnote{https://www.cse.msu.edu/~tangjili/datasetcode/truststudy.htm} is a dataset crawled from the ciaoDVD website which describes user ratings towards movies ranging from $ 1 $ to $ 5 $; 
	(3) \textbf{Epinions}\footnote{https://www.cse.msu.edu/~tangjili/datasetcode/truststudy.htm} dataset records user ratings on different types of items (software, music, television show, etc.) scaled from $ 1 $ to $ 5 $.
	(4) \textbf{WeChat} dataset contains users' clicks on different articles, recorded by the WeChat platform.
	The detailed statistics of the datasets is given in Table~\ref{tab:dataset}.
	Following~\cite{Wang2019graph}, we convert the explicit ratings (ranging from $ 1 $ to $ 5 $) in Last-FM, Ciao, and Epinions dataset into implicit labels where each one is marked as $ 1 $ indicating that user has positive feedback, otherwise, marked as $ 0 $. The threshold for the positive rating is set to be $ 4 $, similar as~\cite{Wang2019graph}. We use MetaPath2vec~\cite{dong2017metapath2vec} to produce the pre-trained embeddings of different nodes in the dataset and feed them into the GCN model as the raw features.
	\begin{table}[htbp]
		\small
		\centering
		\caption{Statistics of datasets.}
		\vskip -1em
		\setlength{\tabcolsep}{4mm}{
			\begin{tabular}{cccc}
				\toprule
				Dataset & \#Users & \#Items & \# Interections \\
				\midrule
				LastFM & 1,892 & 17,632 & 86,769 \\
				Ciao  & 7,375  & 105,114 & 264,229 \\
				Epinions & 22,164 & 296,277 & 857,165 \\
				WeChat & 180,871 & 116,551 & 3,801,612 \\
				\bottomrule
			\end{tabular}%
		}	\label{tab:dataset}
		\vskip -1em
	\end{table}

	\smallskip\noindent\textbf{Evaluation Protocols.}
	We randomly split the entire user-item recommendation records of each dataset into a training set, a validation set, and a test set, where each of them contains 80\%, 10\%, and 10\% of the full records, respectively. 
	Two popular metrics are adopted to evaluate the recommendation accuracy, i.e., 1) the Area Under receiver operator characteristic Curve (AUC) and 2) the Normalized Discounted Cumulative Gain (NDCG). 
	Generally, higher metric values indicate better recommendation accuracy.
	To evaluate NDCG on top-K recommendation performance, we follow a similar setting as~\cite{hu2018leveraging, he2017neural}. Specifically, for each positive item in the test set, we choose $ 50 $ negative items from the set of items which have no interaction records with the target user. Then, we rank the list of positive and negative items together. The final NDCG of each dataset is computed by first averaging over all the test items of a user and then averaging over all the users in the test set. We report the average score at $ N = 10 $ (i.e., NDCG@10) in this paper.

	\smallskip\noindent\textbf{Comparison Methods.}
	We compare four different neighbor sampling methods:
	\textbf{(1)}~Random walk based sampling~\cite{ying2018graph}, which simulates random walks starting from each node and compute the L1-normalized visit count of neighbors visited by the random walk.
	\textbf{(2)}~First-order proximity based sampling~\cite{fan2019meirec, Wang2019graph, wu2019dual, wang2019knowledge}, which examines the neighborhood similarity based on the edge weights~(e.g., number of clicks).
	\textbf{(3)}~Second-order proximity based sampling~\cite{zhao2019intentgc}, which examines the neighborhood similarity based on the number of common neighbors.
	\textbf{(4)}~Our proposed DA similarity based sampling.
	We also compare the following model architectures for node representation: 
	\textbf{(I)}~MEIRec~\cite{fan2019meirec} which is a multi-layer GCN model. MEIRec adopts metapath-guided aggregations to learn user/item representation and samples the neighbors using metapath-based first-order proximity. 
	\textbf{(II)}~IntentGC~\cite{zhao2019intentgc} which is also a multi-layer GCN model. IntentGC learns user/item representation with a faster architecture named IntentNet which avoids unnecessary feature interactions to speed up training.  
	\textbf{(III)}~Our proposed simplified architecture with only one GCN layer.
	Moreover, we extend MEIRec and IntentGC to learn with the DA similarity based sampling method, which are referred to as MEIRec++ and IntentGC++, respectively.
	
	\smallskip\noindent\textbf{Parameter Settings.}
	The optimal parameter settings for all the comparison methods are achieved by either empirical study or suggested settings by the original papers. For all models, we fix the total number of sampled neighbors to be $ 25 $ on all datasets.
	For SLGCN, we adopt Adam~\cite{kingma2014adam} as the optimizer, and set the learning rate as 0.01; the $L_2$ regularization coefficient as $10^{-5}$. We utilize warm-up technique to accelerate the training of the SLGCN. Specifically, we start with an initial batch size of $100$ and then change it to $10240$ after $ 100 $ batches. Note that SLGCN is able to learn with an extra large batch size due to the simplified propagation step. 
	The linear transformation matrix in (\ref{eq:user-representation}) scales as $\mathbf{W}\in\mathbb{R}^{256\times m}$ where $m$ denotes the dimension of the raw feature. The prediction function in (\ref{eq:slgcn}) is a three-layer MLP and the size of each layer is $512$. Code will be released later.
	
	\subsection{Performance Comparison (Q1)}
	Table~\ref{tab:result} reports the performance on the four datasets w.r.t. AUC and NDCG. Overall, our proposed SLGCN consistently achieves the best performance among all four datasets w.r.t. all evaluation metrics. We summarize the major findings as below.
	
	First, the second-order proximity based models (i.e., IntentGC, SLGCN-2ord) achieve a generally better performance than the first-order proximity based models (i.e., MEIRec, SLGCN-1ord), which indicates that comparing the neighborhood structure to find similar neighbors is more reliable than directly comparing the edge weights. Meanwhile, our proposed DA sampling method can help all GCN-based models~(i.e., MEIRec++, IntentGC++) to obtain a general performance enhancement, which verifies that the DA similarity can well-capture the neighbor similarity. 
	
	Second, when fixing the sampling method, our proposed simplified architecture~(i.e., SLGCN-1ord, SLGCN-2ord) can still outperform the corresponding multi-layer GCN architectures~(i.e., MEIRec, IntentGC). The reason is two-fold. First, the simplified architecture still preserves the local averaging operation, which is the main reason why GCN works well~\cite{wu2019sgc,li2018laplacian}. Second, simplifying the multi-layer architecture into a single-layer architecture can largely reduce the difficulty of parameter fitting thus leading to a higher probability of converging to a better local optimal solution. 
	
	\subsection{Learning Efficiency~(Q2)}
	\label{sec:efficiency}
\begin{figure}[t]
		\centering	
		\includegraphics[width=1\columnwidth]{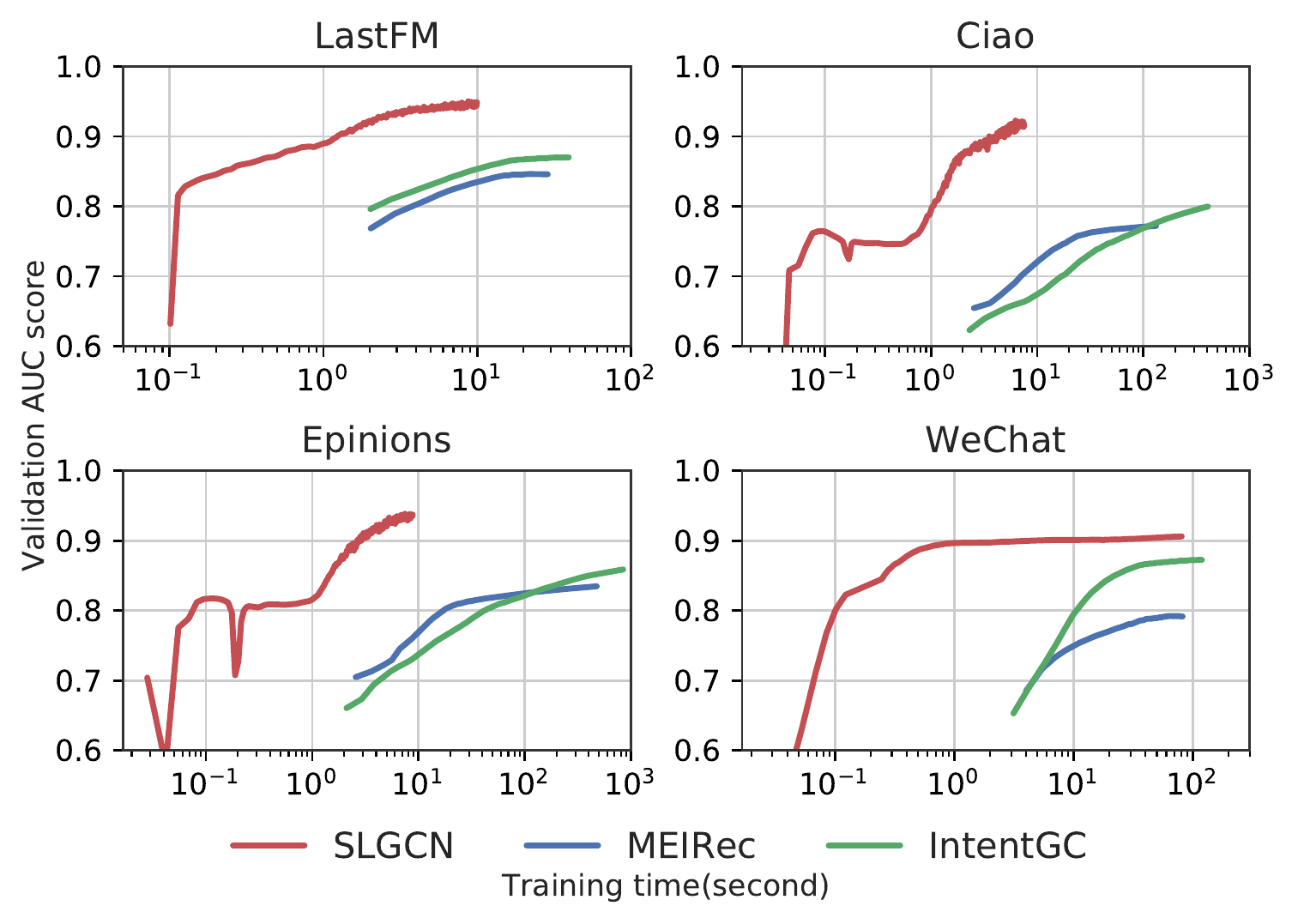}
		\vskip -1em
		\caption{Convergence curves on four datasets~(the horizontal axis is in logarithmic scale). All methods are running on the same GPU device.}
		\vskip -1em
		\label{fig:eff}
	\end{figure}

	One main advantage of SLGCN is the low training complexity. We show the convergence rate and the changes of validation accuracy of all comparing models in Figure~\ref{fig:eff}. Particularly, MEIRec, IntentGC, and SLGCN sample neighbors according to the first-order proximity, the second-order proximity, and the DA similarity, respectively. For all comparing methods, we employ the same mapping function in (\ref{eq:slgcn}) to inference the prediction results and sample the neighbors beforehand so as to present a clean comparison of the training costs.
	All experiments are conducted based on a workstation with $ 24 $ Intel(R) Xeon(R) CPU cores at $ 2.40 $ GHz and one NVIDIA GTX-1080 GPU. The results are averaged over multiple runs. The results in Figure~\ref{fig:eff} shows that SLGCN can achieve superior performance with one or two orders of magnitude speedup in training in all four datasets.
	
	\subsection{Influence From Neighbor Sampling (Q3)}

	Table~\ref{tab:sampling} reports the performance of SLGCN under different sampling methods on four datasets to justify the effectiveness of our propose DA similarity. The results are generated by fixing the model architecture (i.e., node representation and prediction layer) while only varying the neighbor sampling method. 
	
	Overall, the random walk based sampling method generates the worst performance. In fact, the neighbors found by random walks may change significantly when varying the total number or the total length of the generated paths. One can stabilize the results by performing extensive random walks on each node, which, however, is computational exhibitive on large graphs. 
	Following our discussion in~Sec \ref{sec:result}, which mentioned that the DA similarity outperforms the second-order proximity, while the latter outperforms the first-order proximity. 
	This inference can be verified by taking a deeper look at the changes of MANS in Table~\ref{tab:sampling}.
	In particular, we calculate the MANS of user nodes and item nodes separately. The results in Table~\ref{tab:sampling} show that MANS has a general positive correlation with the performance of GCN models. The exceptions are the MANS of user in lastFM and the MANS of item in WeChat, which indicates that we need to assign a lower importance weight $ \lambda_{v,e} $ to the user-click-item similarity in lastFM and the item-click-user similarity in WeChat.

	\subsection{Inference Performance (Q4)}
	\label{sec:architecture}
	\begin{table*}[t]
		\small
		\centering
		\caption{Influence From Neighbor Sampling.}
		\vskip -1em
		\setlength{\textwidth}{2mm}{
			\begin{tabular}{c|ccc|ccc|ccc|ccc}
				\toprule
				& \multicolumn{3}{c|}{LastFM} & \multicolumn{3}{c|}{Ciao} & \multicolumn{3}{c|}{Epinions} & \multicolumn{3}{c}{WeChat} \\
				\midrule
				& \multicolumn{1}{c|}{MANS(U, I)} & \multicolumn{1}{c|}{AUC} & NDCG  & \multicolumn{1}{c|}{MANS(U, I)} & \multicolumn{1}{c|}{AUC} & NDCG  & \multicolumn{1}{c|}{MANS(U, I)} & \multicolumn{1}{c|}{AUC} & NDCG  & \multicolumn{1}{c|}{MANS(U, I)} & \multicolumn{1}{c|}{AUC} & NDCG \\
				\midrule
				rand  & -0.147, -0.653 & 0.9403 & 0.8027 & -0.097, 0.-454 & 0.8828 & 0.7690 & -0.090, -0.445 & 0.9062 & 0.8026 & -0.210, \textbf{-0.502} & 0.8415 & 0.5496 \\
				walk & -0.150, -0.649 & 0.9032 & 0.7698 & -0.104, -0.457 & 0.8599 & 0.7293 & -0.094, -0.446 & 0.8698 & 0.7766 & -0.215, -0.506 & 0.8851 & 0.6907 \\
				1ord  & -0.155, -0.608 & 0.9348 & 0.7856 & -0.096, -0.426 & 0.8656 & 0.7199 & -0.088, -0.431 & 0.9003 & 0.8067 & -0.192, -0.504 & 0.8574 & 0.5550 \\
				2ord  & -0.205, -0.586 & 0.9374 & 0.7871 & -0.094, -0.421 & 0.8929 & 0.7628 & -0.083, -0.411 & 0.9198 & 0.8125 & -0.172, -0.539 & 0.9016 & 0.7411 \\
				\midrule
				sim2  & \textbf{-0.082, -0.356} & \textbf{0.9528} & \textbf{0.8112} & \textbf{-0.082, -0.221} & \textbf{0.9282} & \textbf{0.7957} & \textbf{-0.067, -0.196} & \textbf{0.9403} & \textbf{0.8280} & \textbf{-0.163}, -0.528 & \textbf{0.9104} & \textbf{0.7602} \\
				\bottomrule
			\end{tabular}
		}	\label{tab:sampling}
		\vskip -1em
	\end{table*}

	\begin{figure}[t]
		\centering	
		\includegraphics[width=1\columnwidth]{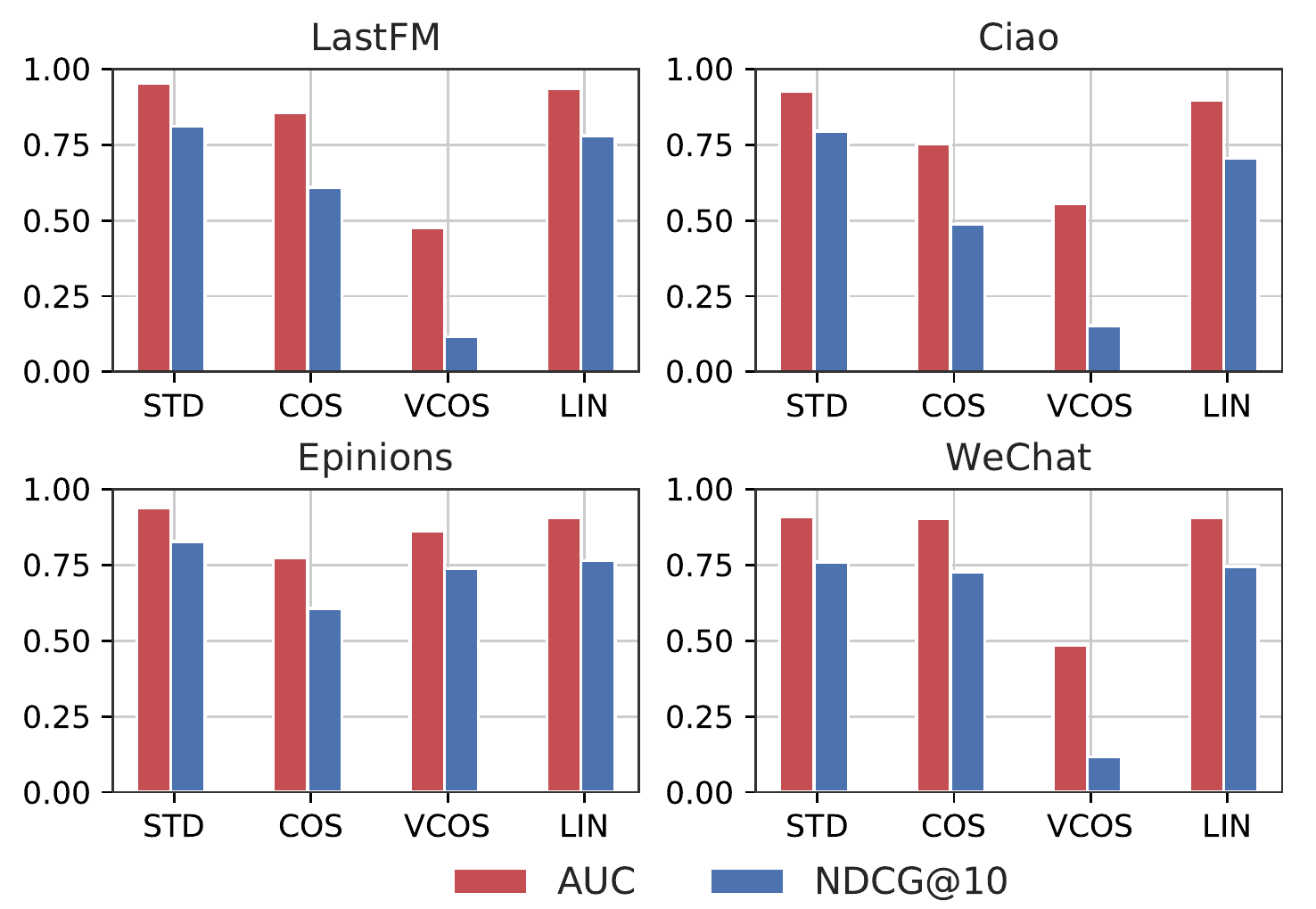}
		\vskip -1em
		\caption{Influence From Model Architecture.}
		\vskip -1em
		\label{fig:architecture}
	\end{figure}
	
	The results in Table~\ref{tab:result} and Figure~\ref{fig:eff} already verified the superiority of using a single GCN layer for node representation. We now focus on the comparison of different inference architectures in SLGCN.
	Specifically, we compare the following variants: 
	1) standard SLGCN, which inferences the results with a stack of multiple MLP layers; 
	2) linear SLGCN, which replace the $\mathbf{H}_u$ and $\mathbf{H}_i$ in~(\ref{eq:slgcn}) with the aggregated $\overline{\mathbf{X}}_{u}$ and $\overline{\mathbf{X}}_{i}$, i.e., do not perform separate nonlinear transformations on the user embedding and the item embedding;
	3) vanilla-cosine SLGCN, which computes the distance between $\overline{\mathbf{X}}_{u}$ and $\overline{\mathbf{X}}_{i}$ with a cosine function to inference the results;
	4) cosine SLGCN, which adds an additional nonlinear activation function outside the cosine function in Vanilla-cosine SLGCN when inferencing the results.
	Figure~\ref{fig:architecture} reports the experimental results, where we refer to the above variants as STD, LIN, VCOS, COS for short.
	The results show that the standard SLGCN achieves the best performance on all four datasets. While the linear SLGCN has an obvious performance degradation, which indicates that it is critical to perform nonlinear transformation on user embedding and item embedding separately  before feeding them into the mapping function.
	Moreover, it is noteworthy that the cosine SLGCN achieves a close performance to the standard SLGCN in WeChat dataset, which indicates that it is promising to replace the MLP layers with cosine function to deliver further complexity reduction when learning from large datasets.
	
	% ============================
	%          Section
	% ============================
	\section{Conclusion}
	In this paper, we introduced the SLGCN model which is able to achieve superior performance along with a few orders of magnitude speedup in training compared with existing models.
	We proved that the proposed DA similarity has a positive correlation with the final performance through both theoretical analysis and empirical simulations. Experimental results revealed that existing GCN models could also make use of the proposed DA similarity metric to improve their performances. 
	Meanwhile, we proposed a simplified GCN architecture which employs a single GCN layer to first aggregate information from the neighbors filtered by DA similarity, and then generates the node representations for inference. 
	Extensive experiments verified the superiority of proposed model on both recommendation performance and training speed.
	We hope our study can inspire more future research activities on building a compact but expressive GCN model for recommendations.
	\bibliographystyle{plain}
	\bibliography{KDD2020}

\begin{thebibliography}{10}

\bibitem{boyd2004convex}
Stephen Boyd and Lieven Vandenberghe.
\newblock {\em Convex optimization}.
\newblock Cambridge university press, 2004.

\bibitem{chami2019hgcn}
Ines Chami, Zhitao Ying, Christopher R{\'e}, and Jure Leskovec.
\newblock Hyperbolic graph convolutional neural networks.
\newblock In {\em NIPS}, pages 4869--4880, 2019.

\bibitem{chen2018fastgcn}
Jie Chen, Tengfei Ma, and Cao Xiao.
\newblock Fast{GCN}: fast learning with graph convolutional networks via
  importance sampling.
\newblock In {\em ICLR}, 2018.

\bibitem{dong2017metapath2vec}
Yuxiao Dong, Nitesh~V Chawla, and Ananthram Swami.
\newblock Metapath2vec: Scalable representation learning for heterogeneous
  networks.
\newblock In {\em KDD}, pages 135--144, 2017.

\bibitem{fan2019meirec}
Shaohua Fan, Junxiong Zhu, Xiaotian Han, Chuan Shi, Linmei Hu, Biyu Ma, and
  Yongliang Li.
\newblock Metapath-guided heterogeneous graph neural network for intent
  recommendation.
\newblock In {\em KDD}, pages 2478--2486, 2019.

\bibitem{fan2019graph}
Wenqi Fan, Yao Ma, Qing Li, Yuan He, Eric Zhao, Jiliang Tang, and Dawei Yin.
\newblock Graph neural networks for social recommendation.
\newblock In {\em WWW}, pages 417--426, 2019.

\bibitem{goyal2018graph}
Palash Goyal and Emilio Ferrara.
\newblock Graph embedding techniques, applications, and performance: A survey.
\newblock {\em Knowledge-Based Systems}, 151:78--94, 2018.

\bibitem{hamilton2017inductive}
Will Hamilton, Zhitao Ying, and Jure Leskovec.
\newblock Inductive representation learning on large graphs.
\newblock In {\em NIPS}, pages 1024--1034, 2017.

\bibitem{hamilton2017representation}
William~L Hamilton, Rex Ying, and Jure Leskovec.
\newblock Representation learning on graphs: Methods and applications.
\newblock {\em IEEE Data Engineering Bulletin}, 2017.

\bibitem{he2017neural}
Xiangnan He, Lizi Liao, Hanwang Zhang, Liqiang Nie, Xia Hu, and Tat-Seng Chua.
\newblock Neural collaborative filtering.
\newblock In {\em WWW}, 2017.

\bibitem{hu2018leveraging}
Binbin Hu, Chuan Shi, Wayne~Xin Zhao, and Philip~S Yu.
\newblock Leveraging meta-path based context for top-{N} recommendation with a
  neural co-attention model.
\newblock In {\em KDD}, pages 1531--1540, 2018.

\bibitem{huang2018asgcn}
Wenbing Huang, Tong Zhang, Yu~Rong, and Junzhou Huang.
\newblock Adaptive sampling towards fast graph representation learning.
\newblock In {\em NIPS}, pages 4558--4567, 2018.

\bibitem{kingma2014adam}
Diederik~P Kingma and Jimmy Ba.
\newblock Adam: A method for stochastic optimization.
\newblock {\em arXiv preprint arXiv:1412.6980}, 2014.

\bibitem{kipf2016semi}
Thomas~N Kipf and Max Welling.
\newblock Semi-supervised classification with graph convolutional networks.
\newblock In {\em ICLR}, 2017.

\bibitem{koren2009matrix}
Yehuda Koren, Robert Bell, and Chris Volinsky.
\newblock Matrix factorization techniques for recommender systems.
\newblock {\em Computer}, 42(8):30--37, 2009.

\bibitem{li2018laplacian}
Qimai Li, Zhichao Han, and Xiao-Ming Wu.
\newblock Deeper insights into graph convolutional networks for semi-supervised
  learning.
\newblock In {\em AAAI}, 2018.

\bibitem{mcpherson2001homophily}
Miller McPherson, Lynn Smith-Lovin, and James~M Cook.
\newblock Birds of a feather: Homophily in social networks.
\newblock {\em Annual Review of Sociology}, 27(1):415--444, 2001.

\bibitem{sarwar2001item}
Badrul Sarwar, George Karypis, Joseph Konstan, and John Riedl.
\newblock Item-based collaborative filtering recommendation algorithms.
\newblock In {\em WWW}, pages 285--295, 2001.

\bibitem{tang2015line}
Jian Tang, Meng Qu, Mingzhe Wang, Ming Zhang, Jun Yan, and Qiaozhu Mei.
\newblock {LINE}: Large-scale information network embedding.
\newblock In {\em WWW}, pages 1067--1077, 2015.

\bibitem{velivckovic2017gat}
Petar Veli{\v{c}}kovi{\'c}, Guillem Cucurull, Arantxa Casanova, Adriana Romero,
  Pietro Lio, and Yoshua Bengio.
\newblock Graph attention networks.
\newblock In {\em ICLR}, 2018.

\bibitem{wang2019knowledge}
Hongwei Wang, Fuzheng Zhang, Mengdi Zhang, Jure Leskovec, Miao Zhao, Wenjie Li,
  and Zhongyuan Wang.
\newblock Knowledge-aware graph neural networks with label smoothness
  regularization for recommender systems.
\newblock In {\em KDD}, pages 968--977, 2019.

\bibitem{Wang2019graph}
Hongwei Wang, Miao Zhao, Xing Xie, Wenjie Li, and Minyi Guo.
\newblock Knowledge graph convolutional networks for recommender systems.
\newblock In {\em WWW}, page 3307–3313, New York, NY, USA, 2019.

\bibitem{wang2019kgat}
Xiang Wang, Xiangnan He, Yixin Cao, Meng Liu, and Tat-Seng Chua.
\newblock {KGAT}: Knowledge graph attention network for recommendation.
\newblock In {\em KDD}, pages 950--958, 2019.

\bibitem{wang2019hgat}
Xiao Wang, Houye Ji, Chuan Shi, Bai Wang, Yanfang Ye, Peng Cui, and Philip~S
  Yu.
\newblock Heterogeneous graph attention network.
\newblock In {\em WWW}, pages 2022--2032, 2019.

\bibitem{wu2019sgc}
Felix Wu, Tianyi Zhang, Amauri Holanda~de Souza~Jr, Christopher Fifty, Tao Yu,
  and Kilian~Q Weinberger.
\newblock Simplifying graph convolutional networks.
\newblock In {\em ICML}, 2019.

\bibitem{wu2019dual}
Qitian Wu, Hengrui Zhang, Xiaofeng Gao, Peng He, Paul Weng, Han Gao, and Guihai
  Chen.
\newblock Dual graph attention networks for deep latent representation of
  multifaceted social effects in recommender systems.
\newblock In {\em WWW}, page 2091–2102, New York, NY, USA, 2019.

\bibitem{ying2018graph}
Rex Ying, Ruining He, Kaifeng Chen, Pong Eksombatchai, William~L Hamilton, and
  Jure Leskovec.
\newblock Graph convolutional neural networks for web-scale recommender
  systems.
\newblock In {\em KDD}, pages 974--983, 2018.

\bibitem{zeng2020graphsaint}
Hanqing Zeng, Hongkuan Zhou, Ajitesh Srivastava, Rajgopal Kannan, and Viktor
  Prasanna.
\newblock Graph{SAINT}: Graph sampling based inductive learning method.
\newblock In {\em ICLR}, 2020.

\bibitem{zhang2018linkprediction}
Muhan Zhang and Yixin Chen.
\newblock Link prediction based on graph neural networks.
\newblock In {\em NIPS}, pages 5165--5175, 2018.

\bibitem{zhang2019deep}
Shuai Zhang, Lina Yao, Aixin Sun, and Yi~Tay.
\newblock Deep learning based recommender system: A survey and new
  perspectives.
\newblock {\em ACM Computing Surveys (CSUR)}, 52(1):1--38, 2019.

\bibitem{zhao2017meta}
Huan Zhao, Quanming Yao, Jianda Li, Yangqiu Song, and Dik~Lun Lee.
\newblock Meta-graph based recommendation fusion over heterogeneous information
  networks.
\newblock In {\em KDD}, pages 635--644, 2017.

\bibitem{zhao2019intentgc}
Jun Zhao, Zhou Zhou, Ziyu Guan, Wei Zhao, Wei Ning, Guang Qiu, and Xiaofei He.
\newblock {IntentGC}: a scalable graph convolution framework fusing
  heterogeneous information for recommendation.
\newblock In {\em KDD}, pages 2347--2357, 2019.

\end{thebibliography}
\end{document}